\documentclass[12pt]{article}
\usepackage{amsmath}
\usepackage{amssymb}
\usepackage{amstext}
\usepackage{amscd}
\usepackage{amsfonts}
\usepackage{graphicx}

\begin{document}

\title{Finite temperature collective modes in a two phase coexistence region of asymmetric nuclear matter
\thanks{\it{This work was partially supported by the CONICET, Argentina.}}}
\author{R. M. Aguirre and A. L. De Paoli.\\ \it{Departamento de Fisica, Facultad de Ciencias Exactas}, \\\it{Universidad Nacional de La Plata,} \\
\it{and IFLP, CCT-La Plata, CONICET. Argentina.}}

\maketitle

\begin{abstract}

The relation between collective modes and the phase transition in
low density nuclear matter is examined. The dispersion relations
for collective modes in a linear approach are evaluated within a
Landau-Fermi liquid scheme by assuming coexisting phases in
thermodynamical equilibrium. Temperature and isospin composition
are taken as relevant parameters. The in-medium nuclear
interaction is taken from a recently proposed density functional
model. We found significative modifications in the energy
spectrum, within certain range of temperatures and isospin
asymmetry,  due to the separation of matter into independent
phases. We conclude that detailed calculations should not neglect
this effect.\\


\end{abstract}
\newpage

\section{Introduction}
A known feature of the low density regime of the nuclear interaction
is that matter undergoes a phase transition of the liquid-gas type.
There is significative evidence of its existence in nuclear
collisions experiments at medium and high energies. It would also
manifest in the crust of neutrons stars, where nuclear correlations
have a determinant effect on the transport properties
and, therefore, in the cooling process of stellar matter. \\
Under the conditions of interest, namely low density and
temperature, matter is essentially composed of protons, neutrons,
and circumstantially leptons. Therefore, we have a strongly
interacting binary system, with more than one conserved charge. A
characteristic aspect of this kind of phase transitions
 is that conserved charges do not
distribute homogeneously among the coexisting phases. This property
give rise to the isospin fractionation observed
in heavy ion collisions. \\

Another interesting phenomenon of the bulk nuclear matter is the
excitation of  collective modes with a definite energy spectrum.
This is a well studied subject because of its multiple consequences.
For example, in astrophysical applications it has been established
that collective modes propagating in neutron
matter modify substantially the neutrino scattering rates \cite{IWAMOTO}.\\
A close relation exists between collective modes and the
liquid-gas nuclear transition, in particular the unstable ones are
the precursor of the thermodynamical instability. There is a great
amount of research concerning on one hand the collective modes and
on the other one the low density phase transition in nuclear matter,
however a few number of investigations deal with the relation between them
\cite{GRECO-COLONNA, PROVIDENCIA-BRITO}. For instance, in
\cite{PROVIDENCIA-BRITO} the spinodal instability and the collective
modes are studied in a neutral mixture of nucleons and electrons,
including both iso-scalar and iso-vector density fluctuations. This
is an appropriate scenario if the system evolves out of equilibrium.
However for characteristic  times large enough, a succession of
coexisting phases must be taken into account.\\
In this letter we aim to study the propagation of collective modes
in a medium composed of different phases in thermodynamical
equilibrium. In contrast to \cite{GRECO-COLONNA, PROVIDENCIA-BRITO}
we have chosen the coexistent phases as the reference state over
which perturbations are applied. We assume the system is well
described in a Fermi liquid scheme, and a linear approximation to
the transport equation is taken. For this purpose we used a model of
the nuclear interaction recently proposed
\cite{BALDO-VIDA헤-SCHUCK}, inspired by the Kohn-Sham density
functional approach. Its concise formulation enables us to include
temperature and isospin composition, and to concentrate on the
physical aspects instead of
calculational complexities.\\
Our conclusions could be related to the evaluation of
cooling rates mediated by neutrinos in proto-neutron stars.\\

\section{The model and its equation of state}

The Density Functional Theory (DFT) seems to be an
appropriate field where different formulations of the in-medium
nuclear interaction could converge. Dissimilar approaches like
Brueckner-Hartree-Fock calculations with free-space two nucleon
potentials, relativistic field models in mean field approximation,
non-relativistic effective forces or in-medium chiral perturbation
theory, provide material for DFT. It has the advantage to yield
accurate results within relatively simple calculations. \\
The Kohn-Sham DFT reduces the description of a complex interacting system
to a simple energy functional resembling that of independent
particles moving in an external potential. This procedure is
justified by the Hohenberg and Kohn theorem \cite{HOHENBERG-KOHN}.\\
Within this framework, it has been proposed recently an energy
functional for finite nuclei \cite{BALDO-VIDA헤-SCHUCK}, which can
be decomposed according to
$E=K+E_{SO}+E_{int}^\infty+E_{int}^{RS}+E_C$. The first and second
terms contain the uncorrelated contributions to the kinetic energy
and the spin-orbit splitting respectively, $E_C$ stands for the
Coulomb energy, and $E_{int}^\infty+E_{int}^{FR}$ represents the
independent nucleons moving in a mean nuclear potential. While
$E_{int}^\infty$ describes the bulk matter behavior, the remaining
term collects finite range effects. Since we are concerned with
infinite homogeneous matter, only kinetic and bulk terms are
retained, in particular we have
\[
E_{int}^\infty=\int d^3 r\left[P_s(x)(1-w^2)+P_n(x) w^2 \right] n
\]
Here $n=n_1+n_2$ is the total baryonic density number, sum of the
proton $n_1$ and neutron $n_2$ densities, $w=(n_2-n_1)/n$ is the
isospin asymmetry fraction, and $P_s(x), \; P_n(x)$ are
interpolating polynomials for symmetric and neutron matter
respectively. They are written in terms of $x=n/n_0$ with $n_0=0.17$
fm$^{-3}$, the number density at saturation,
\begin{eqnarray}
P_s(x)&=&\left \{ \begin{array}{ll} \sum_1^5 b_k^{(s)}\, x^k, \; \;
\; x\leq1
\\
P_s(1)+a_1\, (x-1)+a_2 (x-1)^2, \; \; x>1 \end{array}\right.  \\
P_n(x)&=&\sum_1^5 b_k^{(n)} x^k
\end{eqnarray}

The coefficients $b_k^{(s)}, \, b_k^{(n)},\, a_1$, and $a_2$ can
be consulted in \cite{BALDO-VIDA헤-SCHUCK}. The validity of this
expansion extends up to $x\sim 1.4$. \\
The polynomials have been obtained by adjusting the correlation
part of the energy per particle for symmetric and pure neutron
matter, and then making a quadratic approximation in $w$. A detailed
description  of the formalism can be found in \cite{BALDO-MAIERON}.\\
Within a Fermi liquid approach, the single particle spectra can be
found by a functional derivative $\epsilon_{b}(p)=\delta E/\delta
f_b^0$, with $f_b^0(T,p)=\left[1+\exp(\epsilon_b(p)-\mu_b)/T
\right]^{-1}$ the equilibrium statistical distribution function for
a state of isospin $b=1$ (protons) or $b=2$ (neutrons)
\begin{eqnarray}
\epsilon_b(p)=\frac{p^2}{2\, m}&+&\left[ P_s(x)+x
P'_s(x)\right](1-w^2)+\left[ P_n(x)+x P'_n(x)\right] w^2 +\nonumber \\&& 2 w
(w+I_b) \left[ P_s(x)-P_n(x)\right]\nonumber,
\end{eqnarray}
where $m$ is the degenerate nucleon mass, and $I_b=1 (-1)$ for
protons (neutrons). The prime symbol indicates derivation with respect to $x$.\\

For a given temperature and particle density $n_b$, the
corresponding chemical potential is found by the relation:
\[
n_b= 2\,\int \frac{d^3p}{(2 \pi)^3}\, f_b^0(T,p)
\]
The thermodynamical pressure is given by: $P=\sum_b \mu_b\,
n_b-\mathcal{E}+T \mathcal{S}$, where $\mathcal{E}=E/V$ is the
energy density, and the entropy per unit volume is
\[
\mathcal{S}=-2\,\sum_b \int \frac{d^3p}{(2 \pi)^3}\,\left[f_b^0 \log
f_b^0+\left(1- f_b^0\right) \log \left(1- f_b^0\right)\right]
\]
The numerical factor 2 takes into account the spin degeneracy.\\
A homogeneous system at temperature $T$ and isospin composition
$n_1,\, n_2$ remains thermodynamically stable if the free energy
per unit volume $\cal{F}$ is lower than any linear combination
of energies corresponding to independent phases, satisfying the
conservation laws, i. e.
\begin{equation}
\mathcal{F}(T,n_1,n_2)<\lambda \,
\mathcal{F}(T,n'_1,n'_2)+(1-\lambda)\,
\mathcal{F}(T,n''_1,n''_2),\;\; 0<\lambda<1, \label{STABILITY}
\end{equation}
where $n_k=\lambda \, n'_k+(1-\lambda)\, n''_k, \; {(k=1,2)}$, set up
the conservation requirement for particle and isospin number
\cite{MULLER-SEROT}. \\
Alternatively, the condition (\ref{STABILITY}) can be stated as a
set of two equations \cite{MARGUERON-CHOMAZ}, Det$\mathfrak{F}\geq
0$, Tr$\mathfrak{F}\geq 0$, with $\mathfrak{F}_{ij}=\partial^2
\mathcal{F}/\partial n_i
\partial n_j$. They determine the spinodal region in the phase
diagram.\\
In a phase transition the emerging phases satisfy the Gibbs
condition for equilibrium coexistence:
$\mu_1'(T,n_1',n_2')=\mu_1''(T,n_1'',n_2'')$,
$\;\mu_2'(T,n_1',n_2')=\mu_2''(T,n_1'',n_2'')$,
$\;P(T,n_1',n_2')=P(T,n_1'',n_2'')$, which fix the boundary of the
binodal region. The binodal encloses the spinodal region since
phase separation can occur while the condition (\ref{STABILITY})
is locally satisfied \cite{MULLER-SEROT}.
For a given temperature, the coexistence conditions can be easily
interpreted in terms of isobaric representations of the proton and
neutron chemical potentials as functions of the proton abundance $Y = n_1/n$,
as shown in Fig. 1. The points lying on these curves which are the
vertices of a rectangle correspond to those equilibrium states
which coexist in a phase separation. From these four points, the pair of
neutron-proton chemical potentials on the left (right) correspond to the
low (high) density coexisting phase. As it is explained below, these points
are used to determine the equation of state within the binodal zone.
\\
We have studied the equation of state of the system with variable
isospin fraction at low temperatures, some results are shown in Fig.
2. We have used the Gibbs prescription and the conservation laws to
evaluate the two-phase coexisting region, which produces an almost
linear behavior between the end points corresponding to a single
phase state. The free energy of the system is given by
$\mathcal{F}(T,n_1,n_2)=\lambda \,
\mathcal{F}(T,n'_1,n'_2)+(1-\lambda)\, \mathcal{F}(T,n''_1,n''_2)$,
where $\lambda,n_1',n_2',n_1'',n_2''$ had been determined previously
by the Gibbs construction. It must be pointed out that for symmetric
matter each phase keeps $w=0$ with a constant partial density and,
as the total density increases, only the relative volume fraction
changes. That is, the binodal construction reduces to the standard
Maxwell procedure. In the upper panel of this figure the pressure as
a function of the baryonic density is shown for several isospin
asymmetries and T=2 MeV. The system becomes unstable at very low
density, showing a negative compressibility. The only exception
corresponds to pure neutron matter.  In each case the dashed line
represents the theoretical prediction without phase separation. In
the lower panel the free energy  for T=2 MeV is shown for a selected
set of asymmetry values. As in the previous case, dashed lines
correspond to the unphysical situation. It can be appreciated that
the phase
separation effectively causes a lowering of  $\cal{F}$. \\

For a given temperature $T$ the pair of states satisfying the Gibbs
condition determine a closed curve as the pressure ranges within the
set $0 \leq P \leq P_c(T)$. This curve is the intersection of an
isothermal with the binodal surface, which extends from zero to the
critical temperature $T_c$. We exhibit in Fig. 3 projections of
the binodal into the $P - Y$ plane, corresponding to $T=2,6,$ and
$10$ MeV. The area enclosed by the curve decreases with temperature,
reducing to a point at $T_c$. In our calculations we have found
$T_c\simeq 12.8$ MeV. The full curve can be separated into two
sections with common end points, one having $Y=0.5$ and the other
one with $Y=Y_c$, the proton abundance at the critical pressure
$P_c(T)$. Most of the coexisting phases at a given temperature are
represented by a pair of points, located in different sections. In
such a case lower $Y$ values are associated with the less dense
phase. This behavior is compatible with the isospin distillation
observed in multifragmentation: the most dense phase approaches to
isospin symmetric matter while the other one keeps the neutron excess.\\

For further development, we describe here the Landau parameters of
the model. They are defined in terms of a second variation of the
energy density \cite{BAYMPETH}, namely $F_{B\,B'}=V^2\,\partial^2
\mathcal{E}/\partial f_B\,\partial f_{B'}$. In order to simplify the
notation we have resumed in only one symbol the indices for isospin, spin
and linear momentum, i. e. $B\equiv (b,s,\mathbf{p})$. The Landau's
parameters are defined as the Fourier coefficients of an expansion
in terms of the Legendre polynomials
\[
F^l_{BB'}=(l+1/2)\int_{-1}^1 d\nu \,P_l(\nu)\;V^2\frac{\partial^2
\mathcal{E}(\nu)}{\partial f_B\,\partial f_{B'}},\nonumber
\]
where $\nu=\mathbf{p} \cdot \mathbf{p}'/(p\,p\,')$.\\
Within the energy functional model used here, $\mathcal{E}$ does not
depend on $\nu$, therefore the non-zero components correspond only
to $l=0$, giving
\begin{eqnarray}
F_{B\,B'}^0&=&F_{b\,b'}^0=\frac{2}{n}\,[F_0 + (-1)^b \,G_{b\,b'}]
,\nonumber\\F_0\;\;&=&w^2\,(P_n-P_s)+(1+w^2)\,x\,
P_s'-w^2\,x \, P_n', \nonumber\\
G_{b\,b'}&=&(-1)^{b'}\, (P_n-P_s)
+2\,\delta_{b\,b'}\,w\,[\,P_n-P_s+ x \,(P_n'-P_s'\,)]
\label{LPARAM0}
\end{eqnarray}

\section{Collective modes at finite temperature} 

Collective modes are associated to local density fluctuations that
propagate in the nuclear mean field. These fluctuations are the
effect of small perturbations of the occupation distribution $f_B$
around its equilibrium form $f_B^0$. As a conserved charge the
particle density $n_b$ can be written as a summation over the
level occupation using either $f_B$ or $f_B^0$,
\begin{equation}
n_b=\frac{1}{V}\sum_{s,\,\mathbf{p}} f_B(T)=\frac{1}{V}\sum_{s,\,\mathbf{p}} f_B^0(T)
\end{equation}
The momentum can be regarded as discrete by imposing appropriate
boundary conditions.

Local density fluctuations cause small deviations from the
equilibrium distribution, this perturbation is assumed of periodic
oscillatory nature
\begin{eqnarray}
f_B(T,\mathbf{r},t)&=&f_B^0(T)+\delta
f_B(T,\mathbf{r},t)\nonumber\\&=&f_B^0(T)+
\frac{f_B^{\,0}(T)\,[1-f_B^{\,0}(T)]}{T}\,u_B\,e^{i\,
(\mathbf{q}.\mathbf{r}-\omega\,t)} \label{FLUC}
\end{eqnarray}
Our present ansatz for the perturbation $\delta
f_B(T,\mathbf{r},t)$ ensures that in the limit ${T \rightarrow 0}$
only the levels around the Fermi surface contribute to the zero
mode propagation. The quasiparticle energies change accordingly,
since in the Landau-Fermi liquid model they are obtained  in a
self-consistent manner from the distribution function
\cite{BAYMPETH,OUR}
\begin{equation}
\delta \epsilon_B(T,\mathbf{r},t)=\frac{1}{V}\sum_{B'} F_{B\,B'}\,
\delta f_{B'}(T,\mathbf{r},t)
\end{equation}
The propagation of these perturbations at low temperature is
governed by the Landau's kinetic equation. For temperatures well
below the Fermi energy of the system  $T \ll \epsilon_F$, the
collision term can be neglected \cite{LANDAU}
\begin{equation}
\frac{\partial f_B}{\partial t} + \frac{\partial f_B}
{\partial \mathbf{r}}\,\frac{\partial \epsilon_B}{\partial
\mathbf{p}}-
\frac{\partial f_B}{\partial \mathbf{p}}
\,\frac{\partial \epsilon_B}{\partial \mathbf{r}} = 0 \label{KINET}
\end{equation}
Introducing Eq.(\ref{FLUC}) into Eq.(\ref{KINET}) and keeping only
linear terms in the fluctuations, we obtain
\begin{equation}
\frac{\partial \delta f_B}{\partial t} + \frac{\partial \delta f_B}
{\partial \mathbf{r}}\,\frac{\partial \epsilon_B}{\partial
\mathbf{p}}-
\frac{\partial f_B^0}{\partial \mathbf{p}}
\,\frac{\partial \delta \epsilon_B}{\partial \mathbf{r}} = 0 \label{LKINET}
\end{equation}
which can be further reduced to
\begin{equation}
\left[\omega-\frac{(\mathbf{p}\,.\mathbf{q}\,)}{m}\right]
u_B-\frac{1}{V}\,\frac{(\mathbf{p}\,.\mathbf{q}\,)}{m}\,\sum_{B'} F_{B\,B'}\,\frac{f_{B'}^{\,0}(T)\,[1-f_{B'}^{\,0}(T)]}{T}\,u_{B'}\,=0 \label{CKIN}.
\end{equation}
Since the $F_{B\,B'}$ do not depend on $\mathbf{p}$, see Eqs. (\ref{LPARAM0}), we can write
\begin{equation}
\frac{1}{V}\,\sum_{B'} F_{B\,B'}\,\frac{f_{B'}^{\,0}(T)\,[1-f_{B'}^{\,0}(T)]}{T}\,u_{B'}\,
=\sum_{b'}F_{b\,b'}^0\,w_{b'}\label{SUMA}
\end{equation}
where
\begin{equation}
w_b=\frac{1}{\pi^2\, T}\,\int_0^\infty
dp\,p^2\,f_{b}^{\,0}(T,p)\,[1-f_{b}^{\,0}(T,p)]\,u_{b}(p)
\label{WAMP}\end{equation}
and we have used $u_B=u_b(p)$,
$f_{B}^{\,0}(T)=f_{b}^{\,0}(T,p)$ to show explicitly their
dependence on $p$. Therefore the system (\ref{CKIN}) can be
rewritten in the form
\begin{equation}
w_b+\mathcal{C}_b(T)\,\sum_{b'=1}^2\,F_{b\,b'}^0\,w_{b'}\,=0\;\;\;\;\;(b=1,2)\label{CKIN1}
\end{equation}
with
\begin{equation}
\mathcal{C}_b(T)=\frac{1}{\pi^2\, T}\,\int_0^\infty dp\,p^2\,
f_{b}^{\,0}(T,p)\,[1-f_{b}^{\,0}(T,p)]\,\Omega_{0\,0}(V_z/v_p)
\label{COEF} \end{equation} where $v_p=p/m$, $V_z=\omega/q$ is the phase speed of the collective mode, and
\begin{equation}
\Omega_{0\,0}(s)=\int_{\!-1}^{+1}\frac{dy}{2}\,\frac{y}{y-s}=
1+\frac{s}{2}\ln\left(\frac{s-1}{s+1}\right)
\end{equation}
is the Lindhard function \cite{BAYMPETH}.\\
We shall consider the possibility of damped waves, in
which case the real zero-mode frequency $\omega$ acquires an
imaginary component, namely ${\omega \rightarrow \omega- i\,\eta}$. Introducing ${\delta=\eta/\omega}$,
in the case of slightly damped modes where $0 \leq \delta \ll 1$, we have to leading order
\begin{eqnarray}
\mbox{Re} \; \mathcal{C}_b&\!\!\!\!=&\!\!\!\!\frac{1}{\pi^2\,
T}\,\int_0^\infty\!dp\,p^2\;
f_{b}^{\,0}(T,p)\,[1-f_{b}^{\,0}(T,p)]\,
\left[ \,1+\frac{V_z}{2\, v_p}\,\ln\left| \frac{V_z-v_p}{V_z+v_p}\right|\;\right]\nonumber\\
\mbox{Im}\;\mathcal{C}_b&\!\!\!\!=&\!\!\!\!\frac{\delta}{\pi^2\,
T}\,\int_0^\infty\!dp\,p^2\;
f_{b}^{\,0}(T,p)\,[1-f_{b}^{\,0}(T,p)]\,
\left[\,\frac{V_z}{2\,v_p}\,\ln\left| \frac{V_z-v_p}{V_z+v_p}\right|+\frac{V_z^2}{V_z^2-v_p^2}\,\right]\nonumber\\
&\!\!\!\!+&\!\!\!\!
\frac{\mbox{sgn}(\delta)}{2\,\pi}\,m^2\,V_z\,f_{b}^{\,0}(T,m\,V_z)
\end{eqnarray}
On the other hand, instability modes are characterized by
$\omega=0$ and $\eta < 0$, in which case
\begin{eqnarray}
\mbox{Re}\;
\mathcal{C}_b&\!\!\!\!=&\!\!\!\!\frac{1}{\pi^2\,T}\,\int_0^\infty\!dp\,p^2\;
f_b^{\,0}(T,p)\,[1-f_b^{\,0}(T,p)]\,
\left[ \,1-\frac{\eta}{q\,v_p}\,\arctan\left(q\,v_p/\eta\right)\;\right]\nonumber\\
\mbox{Im}\; \mathcal{C}_b &\!\!\!\!=&\!\!\!\!0
\end{eqnarray}
The proper frequencies are identified with the roots of the
determinant of the system of equations (\ref{CKIN1}), which,
neglecting higher orders in $\mbox{Im}\; \mathcal{C}_b$, reduces to
\begin{eqnarray}
[1+F_{1 1}^0\,\mbox{Re}\,\mathcal{C}_1][1+F_{2 2}^0\,\mbox{Re}\,\mathcal{C}_2]-
{F_{1 2}^0}^2 \; \mbox{Re}\,\mathcal{C}_1\,\mbox{Re}\,\mathcal{C}_2&=&0\nonumber\\
\nonumber\\
F_{1 1}^0\,[1+F_{2 2}^0\,\mbox{Re}\,\mathcal{C}_2]\,\mbox{Im}\,\mathcal{C}_1+ F_{2
2}^0\,[1+F_{1 1}^0\,\mbox{Re}\,\mathcal{C}_1]\,\mbox{Im}\,\mathcal{C}_2&&\nonumber\\
-{F_{12}^0}^2\;\mbox{Im}\,(\mathcal{C}_1\,\mathcal{C}_2)&=&0
\label{EIGEN}
\end{eqnarray}
It is easy to show from Eqs.(\ref{SUMA}) and (\ref{WAMP}), that in
the present model the rate of proton to neutron amplitudes is
momentum independent, namely $u_1/u_2=w_1 \mathcal{C}_2/w_2
\mathcal{C}_1$. As done at zero temperature \cite{OUR}, the isospin
character of the proper modes can be classified as iso-scalar
(iso-vector) for ${\mbox{Re}\,(u_1/u_2)>0\; (<0)}$,
respectively \cite{BARAN}.\\

We have verified that along the phase transition the system is composed of two stable
independent phases. In particular the instability modes have completely disappeared and
only stable eigenmodes are present.
Furthermore, in all the cases considered here, no zero sound modes
are found propagating in the lower density phase of the coexisting
region. The reason
is that $F_{2 2}^0$ takes always negative values in this phase.\\
It is convenient to define an average Fermi velocity $V_F$
\begin{equation}
V_F=\frac{1}{m}\,\sum_{b=1}^2\frac{n_b}{n}\,\sqrt{\frac{\int_0^\infty dp\,p^4\,
f_{b}^{\,0}(T,p)\,[1-f_{b}^{\,0}(T,p)]}{\int_0^\infty dp\,p^2\,
f_{b}^{\,0}(T,p)\,[1-f_{b}^{\,0}(T,p)]}}
\end{equation}
as a reference to estimate deviations from the Fermi surface,
which gives a measure of the validity of the approximations.\\
In Fig. 4 we display the typical low temperature dispersion
relation for different isospin composition. At finite temperature the
proper collective modes arise in pairs. One branch is slightly damped,
meanwhile the other one propagates without dissipation.\\
For the sake of comparison, we also include the unphysical results
obtained by neglecting the phase separation. This situation give
rise to, among others, the instability modes. In such a case we plot
${|\eta|/q}$ instead of $V_z$.\\
For isospin symmetric matter (Fig. 4, $w=0$) the mixed phase extends up to
near $n_0$, and as it was noticed each phase stay at constant partial
density during the transition. As a consequence the collective modes
propagate at constant speed $V_z$. At low densities there are two
branches of iso-vector character, which propagate in the liquid
phase and continue for higher densities, beyond the transition. In
addition, a bivaluated stable iso-scalar mode appears at densities
about ${1.1 n_0}$. For iso-vector as well as for iso-scalar modes,
the branch above $V_F$ corresponds to undamped motion. The
dissipation in the iso-vector wave has an average value of ${\delta
\approx 0.15}$, meanwhile the iso-scalar one has ${\delta \approx
0.09 - 0.19}$. Therefore, damping remains
very small in agreement with our assumptions. \\
If phase separation in symmetric matter is disregarded (dashed
lines), a double iso-vector mode appears at very low density with
small values of $V_z$. It grows monotonously with density and joins
smoothly with the solutions for the mixed stable phase. It has a
moderate damping ${\delta \approx 1}$ at low densities, decreasing
to ${\delta \approx 0.15}$ near $n_0$. At the same time, a pair of
iso-scalar excitations arise at densities below ${0.8 \,n_0}$. The
lower one corresponds to a instability mode reflecting the existence
of the spinodal region. In general the instability modes evolving
out of thermodynamical equilibrium are iso-scalar since both isospin
components are equally affected, in a process which eventually leads
to clusterization \cite{BARAN}. The damped stable scalar mode has a
dissipation coefficient growing from ${\delta \approx 1}$ to $4$ as
the density is increased, mainly because it approaches the unstable configuration.\\
Turning to asymmetric matter (Fig. 4, $w=0.2 - 0.6$), this scene changes
gradually as the density range of coexistence reduces with
increasing $w$. At low and medium densities only iso-vector modes
propagate with almost constant velocity ${V_z \sim 0.3}$. For ${w >
0.5}$ these excitations are strongly suppressed around the medium
density zone.
In fact in the denser coexisting phase, the only one which can
sustain collective motion, the neutron-neutron Landau parameter
$F_{2 2}^0$ decreases and can take negative values when $w$
increases. For example, in the case of $w=0.6$ the parameter $F_{2
2}^0$ vanishes around ${n \approx 0.53 n_0}$, and remains negative
up to ${n \approx 0.81 n_0}$. This explains the suppression of the
density collective modes at medium densities. For supranormal
densities the parameter $F_{2 2}^0$ grows again, and four collective
zero modes reappear. For high isospin asymmetry these four branches
are of mixed character, that is, they change from iso-vector to iso-scalar as the density increases.\\
It must be pointed out that in neutron rich environments such as
$w=0.8$ (not shown here), two iso-vector modes with ${V_z \approx
0.3}$ persist in the very low density regime. This scenario of collective waves propagating in highly asymmetric nuclear matter is expected to hold within the inner crust of neutron stars. The precise structure of this crust is still doubtful, but condensed nuclear droplets immersed in a uniform fluid environment of almost pure neutron matter constitutes a plausible assumption \cite{MARUYAMA,XU}.
Although Coulomb effects are responsible of the existence of the droplets, the mean field properties of the asymmetrical extended fluid are governed by the nuclear forces. In fact, a liquid-gas coexistence of this low density asymmetric nuclear matter environment prevents further instabilities, as it was previously stressed. Moreover, this fluid phase can sustain coherent
density fluctuations, which open a channel for neutrino dispersion.
These collective excitations are similar to the spin-wave propagation investigated in \cite{IWAMOTO}. Therefore it is reasonable to expect a strengthening of the effects discussed there,
in particular a more pronounced reduction of the in-medium mean free path of neutrinos. A further investigation will be presented elsewhere \cite{RMAADP}.\\
If the phase separation is not taken into account (dashed lines), a
stronger suppression of the collective modes is found, mainly because of the more
pronounced decrease of $F_{2 2}^0$ with asymmetry. In this situation
medium and high density excitations are affected, and already
for $w=0.4$ all of them have practically been extinguished. Only the
unstable and stable iso-scalar modes of
sub-saturation densities survive in this case.\\
In general for each damped mode the parameter $\delta$ increases a bit with
growing asymmetry $w$, and this enhancement is more important when
only the unstable one phase is considered. We can conclude that the
binodal phase transition tends to stabilize the density collective
zero modes at low and medium densities.

\section{Conclusions}

We have applied a density functional model of the nuclear
interaction fitted to describe asymmetric nuclear matter
properties, to determine the coexistence regime of the liquid-gas
phase transition at densities below the saturation value $n_0$.
This study has been performed at several temperatures and isospin
asymmetries. The binodal region is diminished as the temperature
increases, till its critical value $T_c \simeq 12.8$ MeV. For a
given temperature this region also decreases with growing
isospin asymmetry.

We have also studied the propagation of zero sound modes at finite
temperature, using a linearized collisionless Landau kinetic
equation. As the method requires a stable reference state which
supports density fluctuations, we have chosen the coexisting phases
in the binodal instead of the unstable spinodal region. We have
found that collective modes are supported only by the denser phase
of the coexistence region, which favors their propagation at low
total densities. Because of this, the phase speed of the collective
excitations remains almost constant in a wide density range.
Furthermore, since in the denser phase the proton to neutron fraction is
higher than in the lighter one, it favors the stability of the
density zero modes in matter with an overall neutron excess. This could have
significative consequences, for example, in
the scattering rates of neutrinos within the proto-neutron star matter. \\




\begin{figure}[h]
\vspace{-1.cm}
\includegraphics[width=0.85\textwidth, height=1.3\textwidth]{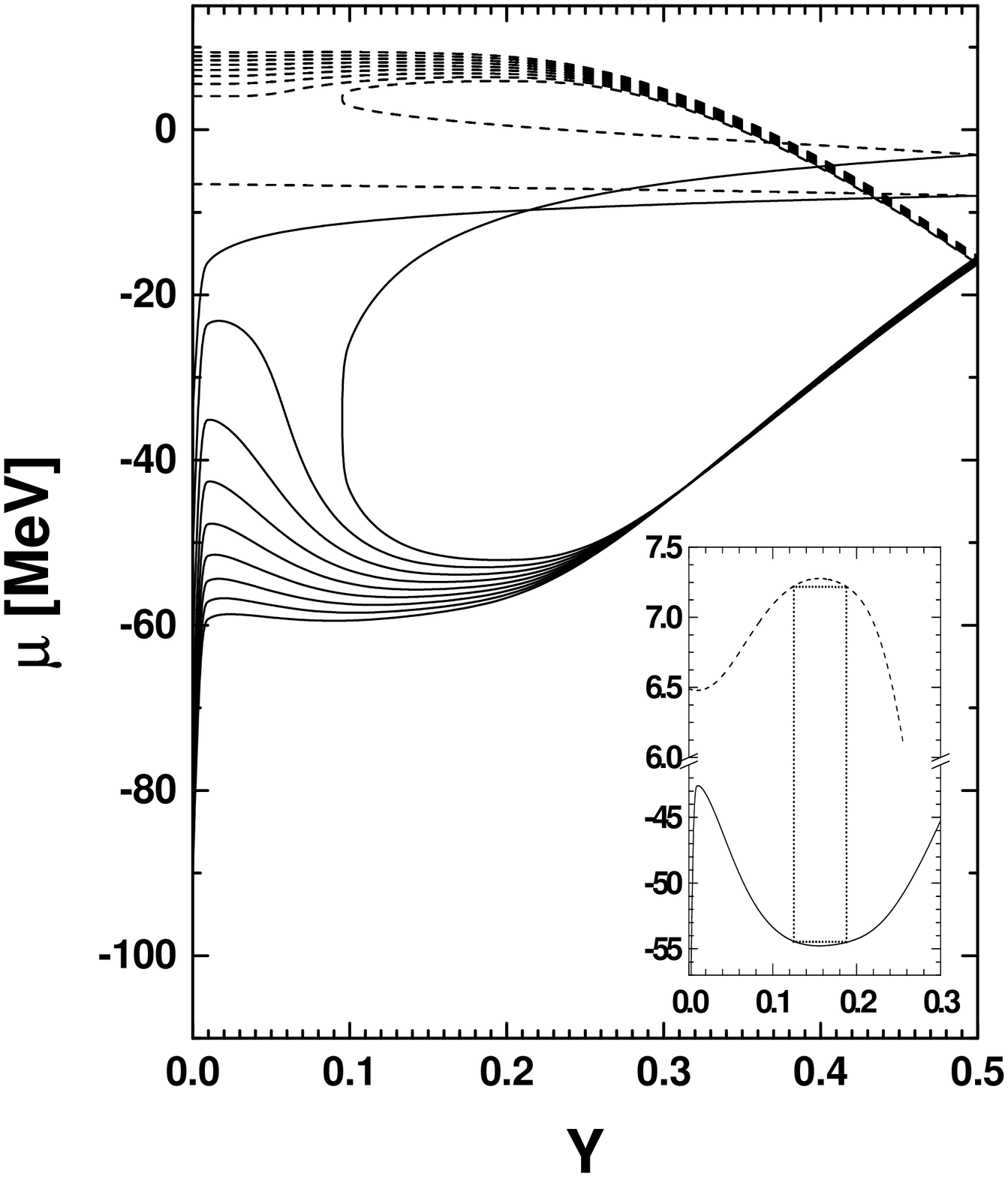} 
\caption{The chemical potential $\mu$ for protons (solid lines) and neutrons (dashed lines) as a function of the proton abundance $Y$ for ${T = 2}$ MeV and a set of fixed pressures $P$, ${0 \leq P \leq 0.19}$ MeV/ fm$^3$. For each isobar both curves for the proton and the neutron chemical potentials are obtained, respectively. The Gibbs construction for a given isobar is shown in the figure inset (dotted line).} \label{FIG1}
\end{figure}

\begin{figure}[h]
\vspace{-1.cm}
\includegraphics[width=1.1\textwidth, height=1.2\textwidth]{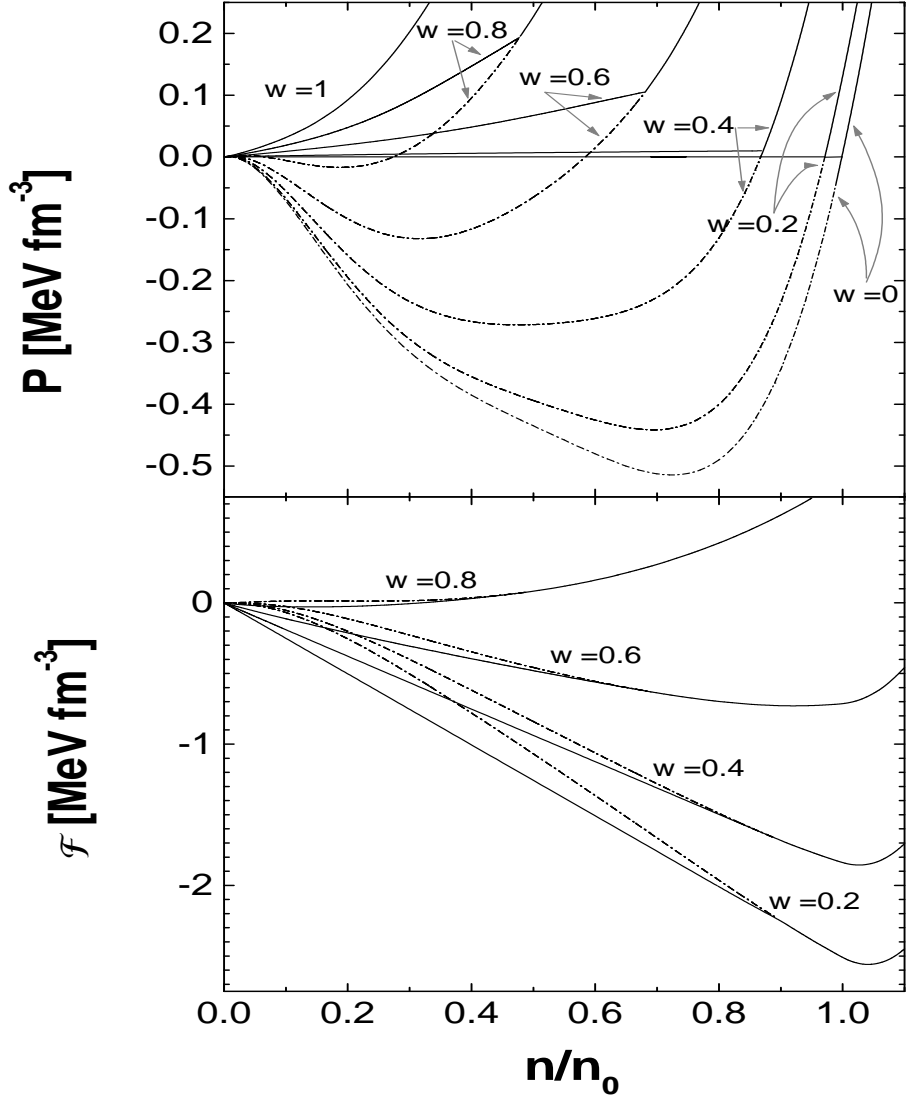} 
\caption{Pressure $P$ and free energy density $\cal{F}$ in terms of the
relative density  $n/n_0$, for ${T = 2}$ MeV and several isospin
asymmetries $w$. Solid lines represent the equilibrium values for
stable matter, the break in the slope of these curves indicates the
end point of the transition. $P$ and $\cal{F}$ evaluated for states out of
equilibrium in the homogeneous one phase are shown by
dashed curves.} \label{FIG2}
\end{figure}

\begin{figure}[h]
\includegraphics[width=0.85\textwidth, height=1.2\textwidth]{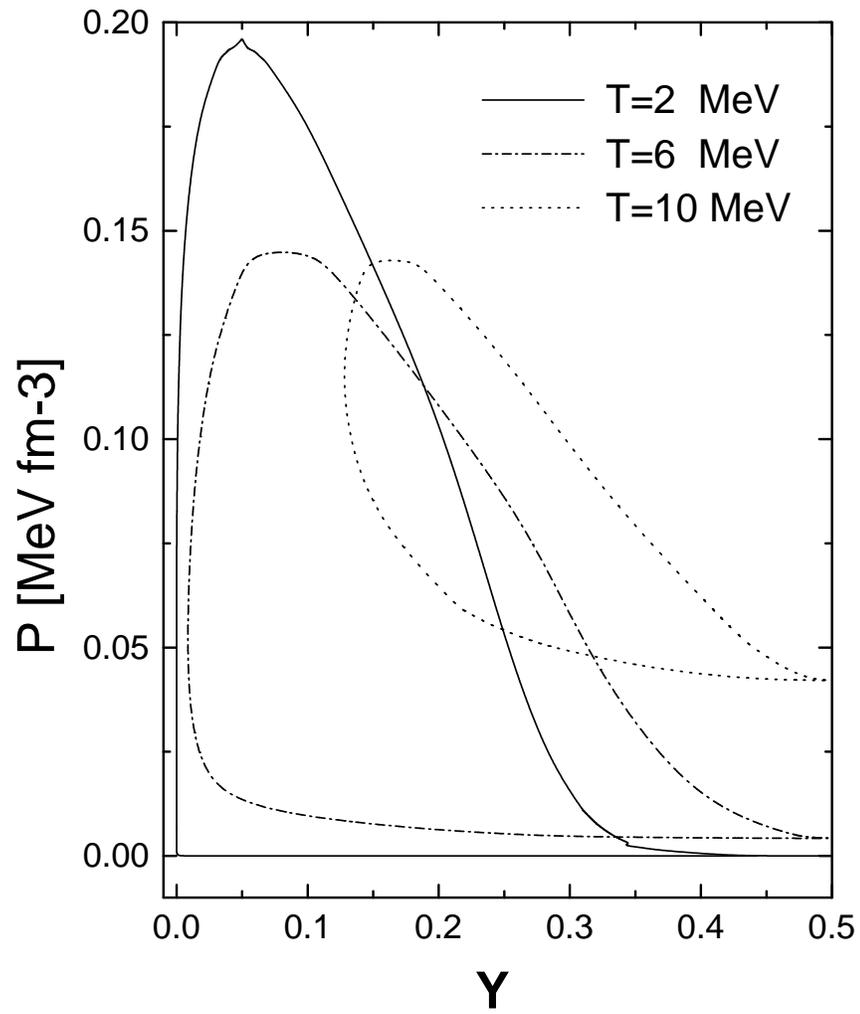} 
\caption{The binodal construction as explained in the text, for
several temperatures. The curves in the ${(P, Y)}$ plane enclose the
zone where two independent phases coexist in thermodynamical
equilibrium.} \label{FIG3}
\end{figure}

\begin{figure}[h]
\centering
\hspace{-1.cm}
\vspace{-0.6cm}
\includegraphics[width=0.9\textwidth, height=1.18\textwidth]{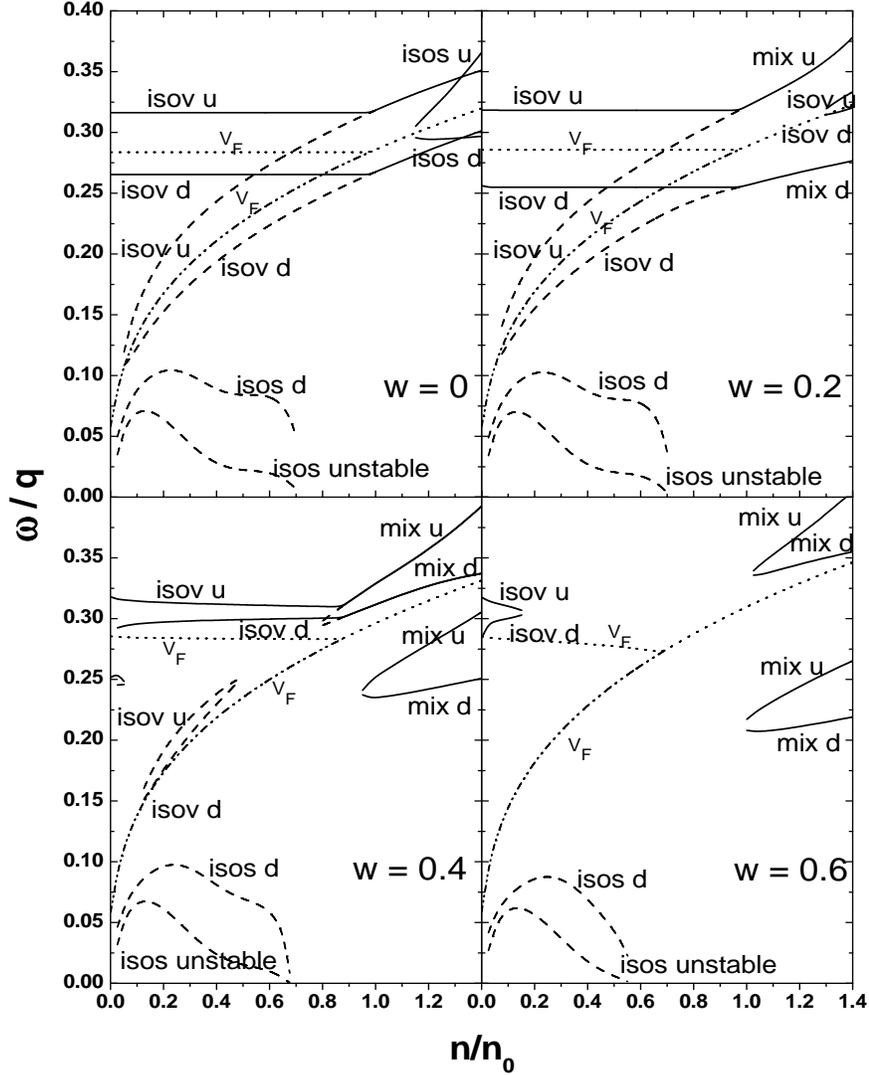}
\caption{Zero sound dispersion $V_z = \omega/q$ at ${T = 1}$ MeV for
several asymmetries $w$ of the nuclear medium. Solid lines
correspond to modes propagating either in the binodal coexisting
phase with the higher density for ${n \lesssim n_0}$, or in the stable
one phase at supra normal densities. Conversely, dashed lines
represent collective propagation within the unstable homogeneous
phase, as explained in the text. The label isos (isov)
indicates the iso-scalar (iso-vector) character, for damped (d) and
undamped (u) stable modes. Curves which change its character from
iso-vector to iso-scalar are labeled mix. For the unstable
iso-scalar curve the ratio ${|\eta|/q}$ is plotted. The average Fermi velocity $V_F$ is shown for the stable high density phase (dot line), and the unstable homogeneous phase (dash-dot-dot line).}
\label{FIG4}
\end{figure}

\end{document}